\documentstyle[12pt,epsfig]{article}

\def\beq{\begin{equation}}
\def\eeq{\end{equation}}

\begin{document}
\title{The production of a non-homogeneous classical pion field and the
distribution of the neutral and charged pions}
\date{}
\author{ A.A. Anselm\ and\ M.G. Ryskin \\
Petersburg Nuclear Physics Institute\\
Gatchina, St.Petersburg 188350, Russia}
\maketitle

\begin{abstract}
The probability distribution $dw/df$ as a function of the ratio
$f=n_0/n_{tot}$ of the neutral to total multiplicities is calculated
for the classical pion fields quickly varying in space and time.
\end{abstract}

1. It is widely discussed in the recent years that the production of a
classical pion field leads to the ratio
\beq
f\ =\ \frac{n_0}{n_{tot}}
\eeq
of the multiplicity of the neutral pions to the total multiplicity
which completely differs from what one expects for the usual mechanisms
of the pion production. For the large values of $n_{tot}$ the standard
mechanisms predicts the probability
\beq
\frac{dw}{df}\ =\ \delta\left(f-\frac13\right)\ ,
\eeq
while for a particular classical state known as the "disoriented chiral
condensate" (DCC), the distribution is
\beq
\frac{dw}{df}\ =\ \frac1{2\sqrt{f}}
\eeq
(see, for example, \cite{1} and a mini-review \cite{2} and the
references therein).

In this letter we shall discuss a class of classical chiral fields
described in \cite{3} and show that the distribution (3) is correct
only for the fields slowly varying in space and time (for instance, for
the DCC which is an exact constant) whereas for the fields which vary
rapidly enough the predicted distribution is quite different (the plot
of the latter distribution is shown in Fig.1).

The classical solutions of the non-linear $\sigma$-model found in
\cite{3} can be represented in the following form. As usually the four
fields $\sigma,\vec\pi$ are constrained by the relation:
\beq
\sigma^2+\vec\pi^2\ =\ f^2_\pi\ .
\eeq
If we denote $\Phi_a=f^{-1}_\pi(\vec\pi,\sigma)$, $a=1,2,3,4$, the
solution is
\beq
\Phi_a\ =\ A_a\cos(\theta(x)+\varphi_a)\ ,
\eeq
where the amplitudes $A_a$ and the phases $\varphi_a$ obey the
equations:
\beq
\sum^4_{a=1} A^2_a=2\ ,\qquad \sum^4_{a=1} A^2_a e^{2i\varphi_a}\ =\ 0\
,   \eeq
while the function $\theta(x)$ satisfies the free wave equation:
\beq
\partial^2\theta(x)\ =\ 0\ .
\eeq
The derivation and the discussion of these solutions see in
ref.\cite{3}.

Let us introduce the complex-4-vector in the chiral $O(4)$ space:
\beq
Z_a\ =\ A_a e^{i\varphi_a}\ =\ X_a+iY_a\ .
\eeq
Eqs. (6) can be rewritten
\beq
X^2_a=1\ , \qquad Y^2_a=1\ , \qquad X_aY_a=0\ .
\eeq
(Here $X^2_a = \sum^4_{a=1}X^2_a$, etc.).

The $O(4)$ invariant dynamics for the production of the states
characterized by the parameters $X_a,Y_a$ can lead to the distribution
function which depends only on the invariants $X^2_a,Y^2_a,X_a\cdot
Y_a$:
\beq dw\ =\ \rho(X^2_a,Y^2_a,X_a\cdot Y_a)\, d^4Xd^4Y\ .  \eeq
Due to
Eqs.(9) the function  $\rho$ is proportional to the $\delta$-functions:
$\delta(X^2_a-1)\delta(Y^2_a-1)\delta(X_aY_a)$, and since all the
invariants turn out to be fixed (10) takes the form:
\beq dw\ =\
\mbox{ const }\delta(X^2_a-1)\delta(Y^2_a-1)\delta(X_aY_a)d^4Xd^4Y\ .
\eeq

The number  of the neutral pions $n_0$ and the total number of the
pions $n_{tot}$ are the functions of $X_a,Y_a$. The distribution in the
variable $f$, the fraction of neutral pions, is given by:
\beq
\frac{dw}{df}\ =\mbox{ const}\int\delta\left(f-\frac{n_0(X,Y)}{n_{tot}
(X,Y)}\right)\delta(X^2_a-1)\delta(Y^2_a-1)\delta(X_aY_a)d^4Xd^4Y.
\eeq

The ratio $n_0/n_{tot}$ is proportional to ratio of the squared
amplitudes of the corresponding pion fields integrated over the
production volume:
\beq
\frac{n_0(X,Y)}{n_{tot}(X,Y)}\ =\ \frac{\int d^3r\ \Phi^2_3}{\int d^3r
[\Phi^2_1+\Phi^2_2+\Phi^2_3]}\ .
\eeq
According to (5):
\beq
\Phi^2_i\ =\ A^2_i\cos^2\bigg(\theta(x)+\varphi_i\bigg)\ ,\  \qquad
i=1,2,3\ .
\eeq
Two different limiting cases are possible.

If $\theta(x)$ varies slowly through the production volume,
$\theta(x)\approx$const, one can redefine the phases and put
$\theta=0$. Then
\begin{eqnarray}
\Phi^2_i & = & A^2_i\cos^2\varphi_i\ =\ X^2_i\ , \nonumber\\
\frac{n_0(X,Y)}{n_{tot}(X,Y)} & = & \frac{X^2_3}{X^2_1+X^2_2+X^2_3}\ .
\end{eqnarray}
This case of the constant field is actually the case of the DCC.

Substituting (15) into Eq.(12) we can easily perform an integration
over $d^4Y$ and the three of the four $d^4X$ integrations to get:
\beq
\frac{dw}{df}\ = \mbox{ const}\int^{+1}_{-1}d(\cos\theta)\delta
(f-(\cos\theta)^2)=\mbox{ const }\frac1{\sqrt{f}}.
\eeq
Here \beq
\cos\theta\ =\ \frac{X_3}{\sqrt{X^2_1+X^2_2+X^2_3}}\ .
\eeq
Thus we see that the inverse square root low comes out when the field
varies slowly through the production volume.

For the opposite case we write $\cos^2(\theta(x)+\varphi_i)=1/2(1+\cos
(2\theta(x)+2\varphi_i))$ and assume that the term linear in
$\cos(2\theta+2\varphi_i)$ is integrated out in Eq.(13). Then
\beq
\frac{n_0(X,Y)}{n_{tot}(X,Y)}=\frac{A^2}{\vec A^2}=\frac{X^2_3+Y^2_3}{
\vec X^2+\vec Y^2}\ , \qquad \vec X^2=\sum^3_{i=1}X^2_i,\ \qquad\vec
Y^2=\sum^3_{i=1}Y^2_i\ .
\eeq

The integral, which appears when this expression is substituted into
Eq.(12), is less trivial then the one for the previous case. In fact it
appears possible to perform the 7 of the 8 integrations $d^4Xd^4Y$
analytically.  This is described in the second part of this letter. A
reader who is not interested in the technical details turn to the plot
of $J(f)=dw/df$ shown in Fig.1. Note also that
\beq
J(0)\ =\ \frac43\ ,\qquad J\left(\frac12\right)\ =\ -\frac5{2\sqrt{2}}
\ln\left|\tan\frac\pi8\right|+\frac12\ =\ 2.058\ , \qquad J(1)\ =\ 0\ .
\eeq
The discontinuity of $dJ(f)/df$ at $f=1/2$ is elucidated qualitatively
in the next section of the letter.\\

2. The properly normalized distribution which emerges from Eqs. (12)
and (8) is
\begin{eqnarray}
\frac{dw}{df}=J(f)&=&\frac1{2\pi^3}\int d^4Xd^4Y\,\delta\left(f -
\frac{X^2_3+Y^2_3}{\vec X^2+\vec Y^2}\right)\delta(x^2_a-1)\delta(
Y^2_a-1)\delta(X_aY_a)\ , \nonumber\\
\int^1_0 J(f)df & = & 1\ .
\end{eqnarray}

We split  $d^4X=d^2x_\perp d^2x_{\|}$ and $d^4Y=d^2y_\perp d^2y_{\|}$,
where $x_\perp=(X_1,X_2)$, $x_{\|}=(X_3,X_4)$, $y_\perp=(Y_1,Y_2)$,
$y_{\|}=(Y_3,Y_4)$, and perform the integrations $d^2x_\perp
d^2y_\perp$ using the three $\delta$-functions. One easily obtains:
\begin{eqnarray}
J(f)&=&\frac1{2\pi^2}\int d^2x_{\|}dy_{\|}\delta\left(f-\frac{X^2_3
+y^2_3}{2-X^2_4-Y^2_4}\right)\ \times \nonumber \\
&\times & \frac{\Theta[(1-x^2_{\|})(1-y^2_{\|})
-(\vec x_{\|}\vec y_{\|})^2]\Theta(1-x^2_{\|})\Theta(1-y^2_{\|})}{
\sqrt{(1-x^2_{\|})(1-y^2_{\|})-(\vec x_{\|}\vec y_{\|})^2}}\ ,
\end{eqnarray}
where $\Theta'$s are the step-functions. The last two $\Theta$
functions can be rewritten in the following way. To have $1-x^2_{\|}>0$
and $1-y^2_{\|}>0$ it is necessary and sufficient to have the product
of these factors and their sum to be positive. The product
$(1-x^2_{\|})(1-y^2_{\|})$ is anyway positive due to the first $\Theta$
function. Therefore we can change $\Theta(1-x^2_{\|})\Theta(1-y^2_{\|})
\to\Theta(2-x^2_{\|}-y^2_{\|})$.

Using Eq.(8) we rewrite the integrations $d^2x_{\|}d^2y_{\|} =A_3dA_3
d\varphi_3A_4dA_4d\varphi_4$ and readily get for $J(f)$:
\beq
J(f)\ =\ \frac4\pi\int^{\pi/2}_0d\phi\int^1_0d\lambda(1-\lambda)
\frac{\Theta[1-2\lambda-2f(1-\lambda)+4\lambda(1-\lambda)f\sin^2\varphi
]}{\sqrt{1-2\lambda-2f(1-\lambda)+4f\lambda(1-\lambda)\sin^2\varphi}}.
\eeq
In Eq.(22) we used $\lambda=1/2\ A^2_4$ and
$\varphi=\varphi_3-\varphi_4$.

Two limiting cases are immediately obtained from (22) . For $f=1$ the
argument of the $\Theta$ function is negative (except for one point:
$\lambda=1/2$, $\sin^2\varphi=1)$ and therefore $J(1)=0$. For $f=0$ one
has:
\beq
J(0)\ =\ 2\int^{1/2}_0 \frac{d\lambda(1-\lambda)}{\sqrt{1-2\lambda}}\ =
\ \frac43\ .
\eeq

For the arbitrary $f$ it is convenient to perform the $\lambda$
integration to get:
\beq
\left.J(f)=\frac4\pi\int\limits^{\pi/2}_0d\varphi\left\{\frac{\sqrt{a
\lambda^2+b\lambda+c}}{-a}-\frac{1+b/2a}{\sqrt{-a}}\arcsin
\frac{2a\lambda+b}{\sqrt{b^2-4ac}}\right\}\right|^{\lambda=
\lambda_{\max},\ {\rm or}\ 1}_{\lambda=\lambda_{\min},\ {\rm or}\ 0}.
\eeq
Here
$$ a=-4f\sin^2\varphi, \qquad b=-2(1-f)+4f\sin^2\varphi, \qquad
c=1-2f\ , $$
\beq
\lambda_{\max,\min}\ =\ -\ \frac b{2a}\ \pm\ \sqrt{
\frac{b^2-4ac}{4a^2}}\ .
\eeq

The limits $(\lambda_{\min},\lambda_{\max})$ in Eq.(24) should be
understood in the following way. If $b^2-4ac<0$ then $J=0$ (since the
argument of the $\Theta(a\lambda^2+b\lambda+c)$ in Eq.(22) is
negative). For $b^2-4ac>0$ the integration in $\lambda$ is taken
between $\lambda_{\min}$ and $\lambda_{\max}$ only if these quantities
are inside the interval (0,1).

The last integration in (24) over $\varphi$ has been performed
numerically and led to the plot for $J(f)$ depicted in Fig.1. We would
like to make two comments concerning this plot.

First, one can analytically calculate $J(f)$ for $f=1/2$. From Eq.(24)
one obtains:
\beq
J(1/2)=\sqrt{2}\int\limits^{\pi/2}_{\pi/4}d\varphi\left[\frac1{\sin
\varphi}+\frac1{2\sin^3\varphi}\right]=-\frac5{2\sqrt{2}}\ln\tan
\frac\pi8+\frac12\ =\ 2.058\ .
\eeq

Second, to understand qualitatively the discontinuity of $dJ(f)/df$ at
$f=1/2$ one can notice that for the fixed value $\varphi=\pi/4$ the
integrand in Eq.(24) has a gap for $f=1/2+\varepsilon$ when
$\varepsilon\to+0$.

Indeed, for $\varphi=\pi/4$ one has $a=-1-2\varepsilon$, $b=
+4\varepsilon$, $c=-2\varepsilon$.
Since it will be immediately evident that only small
$\lambda\sim\sqrt{|\varepsilon|}$ are essential the polynom
$a\lambda^2+b\lambda+c\simeq-\lambda^2-2\varepsilon$. The integral (22) in
$\lambda$ (i.e. the integrand in(24)) is then
\beq
\frac4\pi \int^1_0 d\lambda\ \frac{\Theta(-2\varepsilon -\lambda^2)}{
\sqrt{-2\varepsilon -\lambda^2}}\ =\ \left\{ \begin{array}{lll}
4 & {\rm for } & \varepsilon <0 \\
0 & {\rm for } &\varepsilon >0\ . \end{array} \right.
\eeq

The integration over $\varphi$ (within the interval $\Delta\,\varphi
\sim\sqrt{\varepsilon}$, where $b\lambda\le|c|$ smears out the
discontinuity in $J(f)$ but the derivative $dJ/df$ still has
"singularity" $dJ/df\sim-\Theta(\varepsilon)/\sqrt{\varepsilon}$ at
$\varepsilon\to+0$.\\

3. In ref.\cite{2} we discussed the possible production mechanisms and
the signatures which can be used for the observation of the classical
pion field. One of the conclusions of this paper was that it is not
easy to observe the DCC, or almost constant pion field, due to a simple
reason: the energy density for such a state is so small that no more
than tenths of the pions can be produced through the DCC decay. We
claimed that the observation of the field varying in space and time may
be more promissing since in this case the energy density can be much
higher. We see now that the distribution in $f=n_0/n_{tot}$ for such
fields may be quite different from what is usually supposed, i.e. from
$dw/df\sim1/\sqrt{f}$, and that for a class of the fields described in
\cite{3} $dw/df$ is given by Eq.(24) (Fig.1). It also seems that the
latter fields are closer to those discussed in \cite{4} that the
constant DCC.\footnote{We are grateful to K.Rajagopal for this comment.}

We would like to thank K.Rajagopal and A.Shuvaev for useful
discussions.

The work was supported by the Russian Fund of Fundamental Research
(96-02-18013) and by the INTAS grant 93-0283.

\begin{figure}[h]
\centerline{\epsfig{file=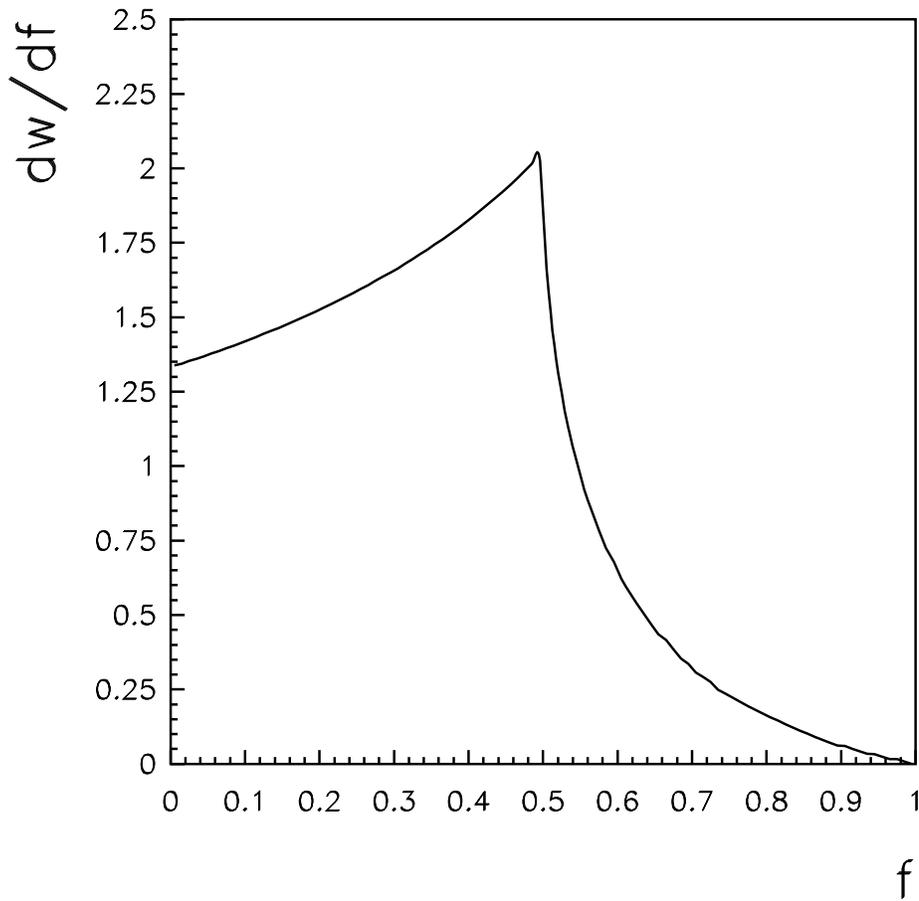,width=13cm}}
\caption{The probability distribution $dw/df$ over the neutral to
total multiplicities ratio $f=n_0/n_{tot}$ for  classical pion fields
quickly varying in space and time.}
\end{figure}

\end{document}